\documentclass[twocolumn,american,pra]{revtex4-1}
\usepackage[T1]{fontenc}
\pdfoutput=1 
\usepackage[latin9]{inputenc}
\usepackage{amsmath}
\usepackage{amssymb}
\usepackage{mathtools}
\usepackage{graphicx}
\usepackage{esint}

\makeatletter
\@ifundefined{textcolor}{}
{%
 \definecolor{BLACK}{gray}{0}
 \definecolor{WHITE}{gray}{1}
 \definecolor{RED}{rgb}{1,0,0}
 \definecolor{GREEN}{rgb}{0,1,0}
 \definecolor{BLUE}{rgb}{0,0,1}
 \definecolor{CYAN}{cmyk}{1,0,0,0}
 \definecolor{MAGENTA}{cmyk}{0,1,0,0}
 \definecolor{YELLOW}{cmyk}{0,0,1,0}
}

\makeatother

\usepackage{babel}

\begin{document}

\title{Discrete-Gauss states and the generation of focussing dark beams}

\author{Albert Ferrando}

\affiliation{Departament d'\`Optica. Interdisciplinary Modeling Group, \emph{InterTech.}
Universitat de Val\`encia. Spain}
\begin{abstract}
Discrete-Gauss states are a new class of gaussian solutions of the
free Schr{\"o}dinger equation owning discrete rotational symmetry. They
are obtained by acting with a discrete deformation operator onto Laguerre-Gauss
modes. We present a general analytical construction of these states
and show the necessary and sufficient condition for them to host embedded
dark beams structures. We unveil the intimate connection between discrete
rotational symmetry, orbital angular momentum, and the generation
of focussing dark beams. The distinguishing features of focussing dark beams are
discussed. The potential applications of Discrete-Gauss states in advanced optical trapping and quantum information processing are also briefly discussed. 
\end{abstract}
\maketitle

\section{Introduction}
In quantum mechanics, gaussian pure states are represented by gaussian
wave functions in position or momentum variables \cite{Schumaker1986a}. Pure gaussian states play an important role in non-relativistic quantum mechanics because  their  evolution is also described by simple gaussian wave packets, which, in addition, minimize the Heisenberg uncertainty relation \cite{Sakurai1994}. In classical optics, the concept of gaussian beam is ubiquitous, since it represents an excellent approximation for the spatial distribution of many realistic light beams in the so-called paraxial approximation \cite{Siegman1986}. Gaussian beams are solutions of the paraxial scalar wave
equation (PWE) for the optical field, formally identical to the linear 2D Schr{\"o}dinger equation (L2DSE) for the wave function in quantum mechanics \cite{Marcuse1982}. For this reason, the free spatial propagation of these beams is described by means of gaussian wave functions with identical properties of gaussian wave packets in quantum mechanics \cite{Marcuse1982}. Besides, the fact that gaussian beams present simple transformation rules under the action of arbitrary optical elements makes them a very convenient tool for the description of  a wide variety of optical systems  \cite{Siegman1986}.

It is also known in quantum mechanics that the 2DLSE admits vortex line solutions with quantized orbital
angular momentum (OAM) presenting nontrivial dynamics \cite{Bialynicki-Birula2000a}. In
classical and quantum optics, solutions of the PWE with well-defined OAM play also an important role \cite{Allen1999a,Yao2011}.
In the free propagation
case, the PWE supports gaussian solutions with well-defined OAM, which
are the optical counterparts of vortex lines in quantum mechanics.
Mathematically, these solutions are given by the so-called Laguerre-Gauss
(\emph{LG}) modes \cite{Siegman1986}. \emph{LG} modes are eigenstates of
the OAM operator and, consequently, also of the $O(2)$ continuous
rotation group operator. They present a phase singularity located
at the axis of symmetry. Since the field intensity is zero at the
singularity, the associated vortex line forms a \emph{dark ray }propagating
in a straight line. The flux around the singularity forms an optical
vortex and it is quantized in such a way the associated topological
charge equals the OAM of the \emph{LG} mode \cite{Allen1999a,Yao2011}.

Nevertheless, in more recent years it has been proven that the L2DSE (from now on, we use this notation to refer also to the PWE) admits more complex solutions with more intricate phase profiles. In this way, multi-singular solutions forming dark rays bundles,
or dark beams, in $(2+1)D$ have been reported in the context
of quantum mechanics \cite{Bialynicki-Birula2000a} and in optics.
In the latter case, a considerable variety of dark beam solutions,
based on  gaussian \emph{LG} modes have been reported \cite{Indebetouw1993,jenkins-joa3_527a,Chavez-Cerda2001a,Bandres2004,Bandres2004a,Volyar2006,Deng2008a,Gutierrez-Vega2008a,Fadeyeva2012a,Steuernagel2012a,Dorilian2013a,Martinez-Castellanos2013a}.
Solutions of the L2DSE with an intricate dark ray structure
forming knots and loops can also be obtained by superposition of \emph{LG}
modes \cite{Berry2001b,Leach2004,Leach2005a,Dennis2010}. Even nontrivial
dark ray solutions of full Maxwell's equations 
can be approximated by superpositions of \emph{LG} modes \cite{Ranada1989a,Bialynicki-Birula2003a,Irvine2008,Kedia2013a}.

Closely related to the appearance of multi-singular solutions is the phenomenon of the breaking of continuous rotational symmetry. The breaking of $O(2)$ symmetry into discrete
rotational symmetry in the \emph{nonlinear} 2D Schr{\"o}dinger equation
is responsible of the so-called vortex transmutation rule \cite{Ferrando2005b,Garcia-March2009},
which is univocally linked to the generation of multi-singular solutions 
\cite{Zacares2009,Garcia-March2009a}. However, the
generation of these multi-singular solutions ---in the form of symmetric  dark rays bundles--- using media with discrete rotational
symmetry has proven to be an essentially \emph{linear} property of the 2D Schr{\"o}dinger equation
\cite{Commeford2012,ferrando2013}. Recently, both the vortex transmutation
rule and the generation of off-axis singularities forming straight dark rays have been experimentally
demonstrated in optics for free linear propagation using discrete
diffractive optical elements (DOE) \cite{Gao2012a,Novoa2014a}. Remarkably, these experimental techniques based on discrete diffractive elements represent a simple form of generation of multi-singular gaussian beams.

Multi-singular solutions, particularly those forming part of a gaussian beam, are excellent candidates for applications in optical trapping. This is so because it is known since long ago that the momentum of light can be used for the acceleration, trapping and levitation of particles by means of radiation pressure \cite{Ashkin1970a,Ashkin1971a,Ashkin2000a}. Gradient forces generated by single-beams exhibiting  adequate intensity gradients constitute the physical mechanism on which optical tweezers are based \cite{Ashkin1986a,Ashkin1992a}. Furthermore, it has been recently proven  that not only intensity but phase gradients can provide useful optical forces for optical trapping, including the transfer of the OAM of light to particles \cite{Grier2003a}. Moreover, optical forces arising from phase gradients can be used complementarily to intensity-gradients traps to taylor force profiles for improved optical trapping \cite{Roichman2008a}. In this way, the control of the properties of both the phase and the intensity profiles of an optical beam turns out to be an essential ingredient for advanced optical trapping \cite{Woerdemann2013a}. The possibility of generating optical beams with a controllable and rich phase and intensity structure is then a source for potential applications in this field. Easily manipulable multi-singular gaussian beams can thus play an important role in this context.  

Multi-singular gaussian states with an embedded nontrivial dark beam structure can also play a relevant and complementary role in quantum information experiments. OAM states of light, physically implemented as gaussian \emph{LG} modes, has been proposed to realize high-dimensional quantum spaces for quantum information applications \cite{Molina-Terriza2007b}. In a similar manner to OAM states, other basis of the Hilbert space of the solutions of the 2DLSE can be also used for similar purposes. In this context, recent experiments in which photons associated to a particular type of multi-singular gaussian modes (Ince-Gauss modes) have been entangled  demonstrate the feasibility and the potentiality of this approach \cite{Krenn2013a}.

In this letter we show an explicit construction of multi-singular gaussian solutions
of the 2DLSE owning discrete rotational symmetry of \emph{any} order, the so-called, Discrete-Gauss (\emph{DG}) states. The construction of these new \emph{DG} states is general and systematic thus permitting to
unequivocally elucidate the intimate relation between discrete rotational
symmetry and the generation of multi-singular solutions. We shall show the necessary and sufficient conditions to generate multi-singular gaussian solutions by the action of an operator that breaks continuous rotational symmetry and which can be easily implemented using diffractive optical elements. Likewise, we will analyze the nature of the rich focussing dark beams structures, i.e., symmetric
dark rays bundles with a focussing point, which can be embedded within \emph{DG} states. Finally, we shall also demonstrate that this set constitutes a basis of the Hilbert space of solutions of the 2DLSE by showing that they verify a biorthogonal relation, completely analogous to that fulfilled by \emph{LG} modes.

\section{The discrete deformation operator}

We start by writing the free 2DLSE in complex variables ($w=x+iy$,
$\overline{w}=x-iy$) and use $\tau$ as a evolution parameter:
\begin{equation}
-i\frac{\partial\phi}{\partial\tau}+\frac{\partial^{2}\phi}{\partial w\partial\overline{w}}=0.
\label{eq:complex_LSE}
\end{equation}
In optics, $\tau=\lambda z$/$\pi$, $\lambda$ being the wavelength
of light and $z$ the axial coordinate. In quantum mechanics, $\tau=\left(2\hbar/M\right)t$
, where $M$ is the particle mass, $\hbar$ is the Planck's constant
and $t$ is time. The equation above admits a rotationally symmetric
gaussian wave packet solution \cite{Siegman1986}:
\begin{equation}
\phi_{00}(w,\overline{w},\tau)=\frac{i\tau_{R}}{q(\tau)}\exp\left(-\frac{iw\overline{w}}{q(\tau)}\right),
\label{eq:fundamental_mode}
\end{equation}
where $q(\tau)=\tau+i\tau_{R}$ (in optics $ $$\tau_{R}=\lambda z_{R}$/$\pi$,
where $z_{R}$ is the Rayleigh length). The complex-argument ``elegant''
Laguerre-Gauss (\emph{LG}) modes can be constructed by reiteratively
applying the differential operators $\partial/\partial w$ and $\partial/\partial\overline{w}$
to the fundamental solution $\phi_{00}$ \cite{Zauderer1986,Enderlein2004a}.
\emph{LG} modes are eigenfunctions of the 3rd component of the orbital
angular momentum operator (OAM), $\hat{L}_{z}\phi_{lp}^{LG}=l\phi_{lp}^{LG}$.
\emph{LG} modes with $l\neq0$ present a single phase singularity
at the origin, where the axis of rotational symmetry is also located.
The topological charge of this single phase singularity ($q=(2\pi)^{-1}\oint d\mathbf{l}\cdot\nabla\arg\phi_{lp}^{LG}$
calculated along a circuit enclosing it) equals its OAM, $q=l$. The
value of the field at the $\tau=0$ plane determines completely the
solution for every value of $\tau$. For \emph{LG} modes, this plane is
particularly characteristic since all modes reach here, like the generating
$\phi_{00}$ function, their minimal width (in optics, this plane
is known as the waist). In addition, at $\tau=0$ \emph{LG} modes with
$p=0$ take the simple form of a gaussian vortex $\phi_{l0}^{LG}(0)\sim\Omega_{w}^{l}\phi_{00}(0)$,
where $\Omega_{w}^{l}\equiv\{w^{|l|}\,(l>0),\overline{w}^{|l|}\,(l<0)\}$.
The gaussian vortex is obtained thus by simply multiplying $\phi_{00}(0)$ by a transfer function $t$, given in this case by $\Omega_{w}^{l}$.

We analyze now a related but different
problem. The $\tau=0$ condition can be chosen differently to change
the rotational symmetry properties of the solution. This can be done
by properly selecting the $t$ function. Let us consider the most general
case given by a condition of the form:
\begin{equation}
\phi(w,\overline{w},0)=t(w,\overline{w})\phi_{00}(w,\overline{w},0),\label{eq:general_condition_at_waist}
\end{equation}
in which now $t(w,\overline{w})$ is an arbitrary non-singular analytical
function in $w$ and $\overline{w}$ .
We write this function as $t=a\exp iV$,
where $a$ is an \emph{arbitrary} complex function and $V$ is a real
analytical function. The $\tau=0$ condition is then 
\begin{equation}
\phi(0)=\exp iV(w,\overline{w}) \psi(0), 
\label{waist_condition}
\end{equation}
where $\psi=a\phi_{00}$ is now an arbitrary function. Although we are
finding solutions of the Schr{\"o}dinger equation with no potential, the
particular form of the $\tau=0$ condition is equivalent to the action
of the quasi-instantaneous potential $V$ at $\tau=0$ on the wave
function $\psi$ \cite{ferrando2013}.  We consider then a real local potential owning
purely discrete rotational symmetry of order $N$ with respect to
the origin (i.e., invariance under the $\mathcal{C}_{N}$ point symmetry
group):
\begin{equation}
V(w,\overline{w})=v\left(w^{N}+\overline{w}^{N}\right),
\label{discrete_potential}
\end{equation}
where $v=\varepsilon\nu$, $\varepsilon$ being a small interval $\tau=\varepsilon>0$
indicating the extension of the quasi-instantaneous action of the
potential in the $\tau$ domain, whereas $\nu$ is the leading order
symmetry parameter characterizing the $N_{\mathrm{th}}$-fold symmetry
of the local potential $V$. Under a transformation $G_{N}$ of this
group, complex coordinates change as: 
\begin{eqnarray}
w & \overset{G_{N}} {\rightarrow}& \epsilon_N w \nonumber \\
\overline{w} & \overset{G_{N}}{\rightarrow} & \epsilon_N^{*} \overline{w},
\label{complex_rotation}
\end{eqnarray}
where $\epsilon_{N}=\exp\left(2\pi i/N\right)$ is the complex elementary
finite rotation of $N_{\mathrm{th}}$ order. Thus $w^{N}$ and $\overline{w}^{N}$
are $\mathcal{C}_{N}$-invariants and, therefore, $V$ is an explicitly invariant potential. The potential $V$ is real if
we choose $v$ to be real, as we shall do. Moreover, $V$ is the most general form of a local expansion of a purely 
invariant $\mathcal{C}_{N}$ real potential around $w=0$. It is easy to prove that the alternative combination 
$v w^{N}+ v^{*} \overline{w}^{N}$ with complex $v$ is equivalent to  Eq.(\ref{discrete_potential}) up to a global rotation.

We consider now that the function $\psi$ in Eq.(\ref{waist_condition}) is a gaussian vortex, i.e.
an elegant \emph{LG} mode with $p=0$, and use $V$ as a way to change
the rotational properties of the solution at the waist of this mode.
For this reason, we define a new modified function at $\tau=0$ given
by:
\begin{eqnarray}
\phi{}_{lNv}(w,\overline{w},0) &= & e^{iV(w,\overline{w})}\phi_{l0}^{LG}(w,\overline{w},0) \nonumber \\
			& \overset{O(v^{2})}{\sim} & \left[1+iv\left(w^{N}+\overline{w}^{N}\right)\right]\phi_{l0}^{LG}(w,\overline{w},0).\nonumber \\
\label{waist_condition_discrete}
\end{eqnarray}
This function gives the first order correction in the small interval
$\tau=\varepsilon$ for the output field scattered by the quasi-instantaneous
discrete potential $V$. A solution of the 2DLSE (\ref{eq:complex_LSE})
fulfilling the condition (\ref{waist_condition_discrete}) is
obtained simply by formally replacing $w$ and $\overline{w}$ by
the differential operators:
\begin{eqnarray}
 \hat{l}_{+} & \equiv & \hat{w}+\tau\hat{\overline{p}} \nonumber \\
 \hat{l}_{-} & \equiv & \hat{\overline{w}}+\tau\hat{p},
\label{l_operators}
\end{eqnarray}
where $\hat{w}$ and $\hat{\overline{w}}$ are the complex
position operators and $\hat{p}=-i\partial/\partial w$ and $\hat{\overline{p}}=-i\partial/\partial\overline{w}$
are their corresponding momentum ones. These operators belong to the Lie algebra of symmetries of the free $(2+1)D$ Schr\"odinger equation \cite{Miller2012a,Wunsche1989}. In this way, the modified field by the presence of the discrete potential takes the form:
\begin{eqnarray}
\negthickspace\negthickspace\negthickspace\phi_{_{lNv}}^{DG}(w,\overline{w},\tau) & \overset{O(v^{2})}{\sim} & \left[1+iv\left(\hat{l}_{+}^{N}+\hat{l}_{-}^{N}\right)\right]\phi_{l0}^{LG}(w,\overline{w},\tau),\label{eq:DG_state}
\end{eqnarray}
 The previous replacement rule can be proven as follows. Since complex
position and momentum operators fulfill standard commutation relations:
$\left[\hat{w},\hat{p}\right]=i=\left[\hat{\overline{w}},\hat{\overline{p}}\right]$,
the commutation relations with respect the evolution operator of equation
(\ref{eq:complex_LSE}) $U(\tau)=\exp\left(i\tau\hat{H}\right)$,
where $\hat{H}=\hat{p}\hat{\overline{p}}$, are
\[
\left[\hat{w},U(\tau)\right]=-\tau\hat{\overline{p}}\, U(\tau),\,\,\,\,\,\,\left[\hat{\overline{w}},U(\tau)\right]=-\tau\hat{p}\, U(\tau).
\]
These relations determine that the position operators $\hat{w}$ and
$\hat{\overline{w}}$ transform into $\hat{l}_{+}$ and $\hat{l}_{-}$,
respectively, under the action of the evolution operator. So, we have
\begin{equation}
U(\tau)\hat{w}=\hat{l}_{+}U(\tau),\,\,\,\,\, U(\tau)\hat{\overline{w}}=\hat{l}_{-}U(\tau).
\label{replacement_rule}
\end{equation}
This property justifies the step from Eq.(\ref{waist_condition_discrete})
to Eq.(\ref{eq:DG_state}) since $\phi_{_{lNv}}^{DG}(\tau)=U(\tau)\phi{}_{lNv}(0)$.

We can see the important role played by the operator
\begin{equation}
\hat{D}_{v}(N)\equiv\exp\left[iv\left(\hat{l}_{+}^{N}+\hat{l}_{-}^{N}\right)\right],
\label{discrete_deformation_operator}
\end{equation}
which transforms an \emph{LG} mode into a new gaussian state, which  in \emph{bra-ket} notation can be written as
\begin{equation}
\left|DG(l,p,N)\right\rangle _{v}=\hat{D}_{v}(N)\left|LG(l,p)\right\rangle.
\label{braket_deformation_equation}
\end{equation}
Inasmuch as $\left|LG(l,p)\right\rangle$ is a solution of the free L2DSE and $U(\tau)e^{iV}=\hat{D}_{v} U(\tau)$, it is immediate to show that 
the new state $\left|DG(l,N,p)\right\rangle _{v}$ also verifies this equation. 

According to the complex transformations (\ref{complex_rotation}) and to their definition (\ref{l_operators}), the operators $\hat{l}_{\pm}$
transform tensorially under  $\mathcal{C}_{N}$ rotations  as: $\hat{l}_{+}\overset{G_{N}}{\rightarrow}\epsilon_{N}\hat{l}_{+}$
and $\hat{l}_{-}\overset{G_{N}}{\rightarrow}\epsilon_{N}^{*}\hat{l}_{-}$, where  $\epsilon_{N}=\exp i 2\pi/N$.
As a consequence, $\hat{D}_{v}$ is an invariant operator
under the $\mathcal{C}_{N}$ group. Moreover, the operator $\hat{D}_{v}$ is \emph{unitary}: $\hat{D}_{v}^{\dagger}\hat{D}_{v}=1$. This is a consequence of the relation $\hat{l}_{+}=\hat{l}_{-}^{\dagger}$, which makes the operator  $v\left(\hat{l}_{+}^{N}+\hat{l}_{-}^{N}\right)$ self-adjoint for $v$ real. Since $\hat{D}_{v}$ changes the rotational  symmetry properties of the \emph{LG} mode from continuous to discrete, we refer to it as the \emph{discrete deformation} operator. 

\section{Discrete-Gauss states}

The states generated by the discrete deformation operator  $\hat{D}_{v}$ by means of the deformation equation (\ref{braket_deformation_equation}) are \emph{Discrete-Gauss states}. We recognize that  Eq.(\ref{eq:DG_state}) is, up to  $O(v^{2})$ terms, nothing but the deformation equation (\ref{braket_deformation_equation}) for
$p=0$ states.
Consequently, $\phi_{_{lNv}}^{DG}$ is
a Discrete-Gauss (\emph{DG}) state with $p=0$. Although it is not strictly necessary from the formal point of view, we will restrict ourselves from now on to discrete deformations of \emph{LG} modes with $p=0$, so the index $p$ will be ignored unless explicitly mentioned.  
\begin{figure}
\includegraphics[width=1\columnwidth]{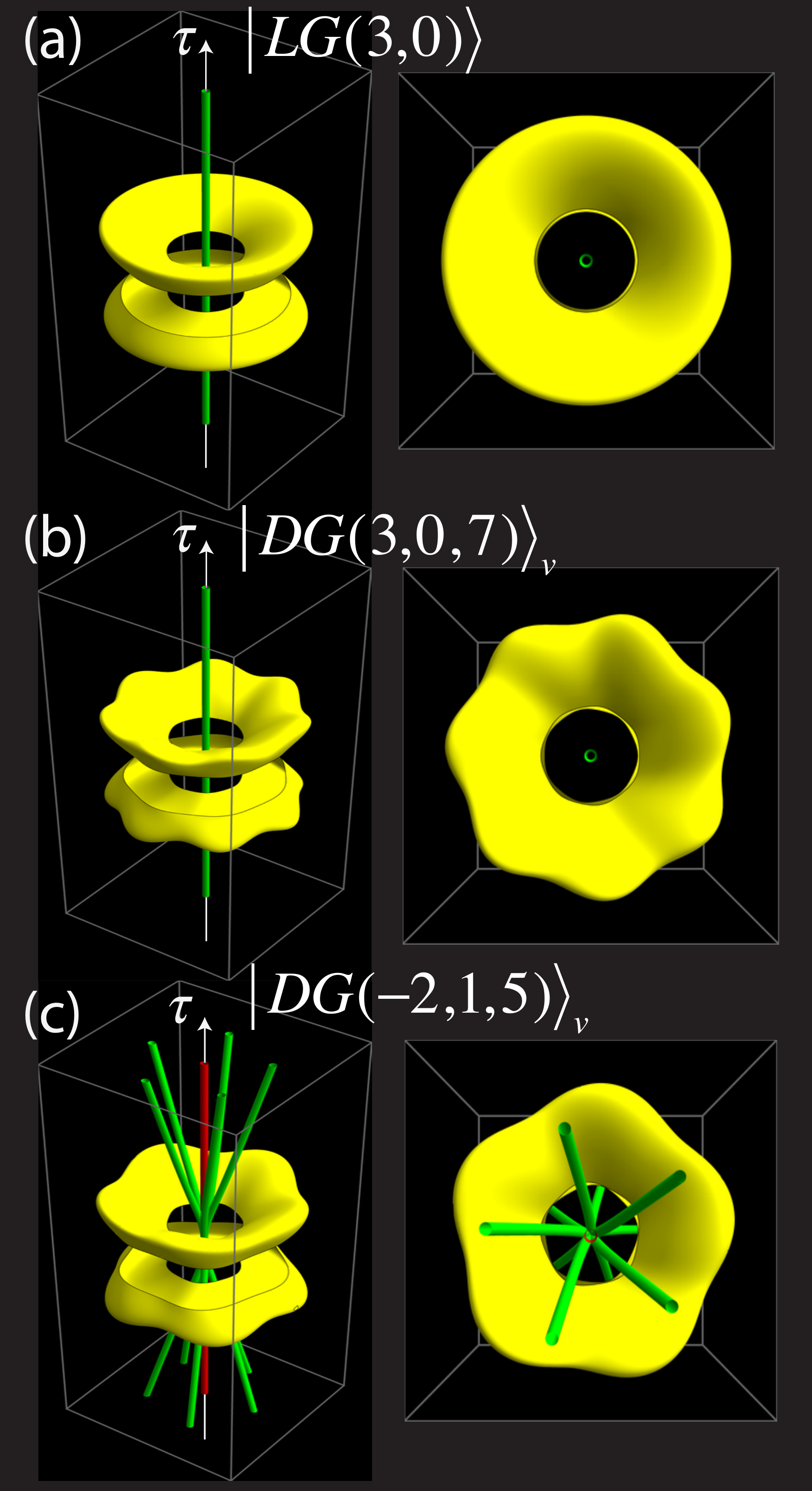}
\caption{Amplitude (3D surface at half-maximum) and singularity structure of
a Laguerre-Gauss mode $LG(l,0)$ and two of its discrete deformations
$DG(m,k,N)$ in the $xy\tau$ space {[}right column, top view; green
(red) color indicates positive (negative) $q$; $\tau_{R}=1$; $v=0.1${]}:
(a) $LG(3,0)$ mode; single dark ray trajectory is shown. (b) \emph{DG}
state with $k=0$ and its single associated dark ray with $q_{\mathrm{ax}}=+3$.
(c) \emph{DG} state with $k=1$ generating a \emph{focussing} dark beam
with $q_{\mathrm{ax}}=-2$ and $N=5$ off-axis $q=+1$ singularities.}
\label{fig1}
\end{figure}
By construction, the rotational symmetry group of a $\left|DG(l,N)\right\rangle _{v}$
is no longer $O(2)$ but $\mathcal{C}_{N}$. Examples of two \emph{DG}
states  $O(v^{2})$ corresponding to two different discrete deformations of a \emph{LG}
mode, which clearly reflect this feature, are given in Fig.\ref{fig1}.  According to Eq.(\ref{braket_deformation_equation}),
$\left|DG(l,N)\right\rangle _{v}$ states are eigenfunctions of the
\emph{discrete} rotation operator $\hat{G}_{N}$: 
\begin{equation}
\hat{G}_{N}\left|DG(l,N)\right\rangle _{v}=\epsilon_{N}^{l}\left|DG(l,N)\right\rangle _{v},
\end{equation}
where  $\epsilon_{N}=\exp i 2\pi/N$. However, they are not eigenfunctions of the OAM operator $\hat{L}_{z}$.
Nevertheless, despite $l$ is not the OAM of the state, it is still
a good quantum number for a \emph{DG} state. Now, it is interpreted as
the ``unfolded'' value of the discrete angular momentum (or angular
pseudo-momentum) $m$, which is the real conserved quantity associated
to the $\mathcal{C}_{N}$ symmetry of the \emph{DG} state \cite{Garcia-March2009}.
Discrete angular momentum $m$ is the corresponding folded value of
$l$ since is bounded $\left|m\right|\le N/2$ and $l=m+kN$, where
$k\in\mathbb{Z}$. The relation between $l$ and $m$ is forced by
the uniqueness of the $\hat{G}_{N}$ eigenvalues $\epsilon_{N}^{l}=\exp\left(2\pi il/N\right)=\epsilon_{N}^{m}$.
This feature is an important characteristic of \emph{DG} states since it clearly unveils the relation between  $l$ (OAM) and $m$ (discrete angular momentum), i.e., between the two conserved quantities associated to spatial rotations in the $O(2)$ (continuous) and  $\mathcal{C}_{N}$ (discrete) cases.

An \emph{LG} mode with $p=0$ takes the form of a gaussian vortex at $\tau=0$, so that $\phi_{l0}^{LG}(0)\sim\Omega_{w}^{l}\phi_{00}(0)$,
where $\Omega_{w}^{l}\equiv\{w^{|l|}\,(l>0),\overline{w}^{|l|}\,(l<0)\}$. Using the replacement rule (\ref{replacement_rule}), this implies that its value for $\tau \ne 0$ can be obtained simply by replacing the complex function $\Omega_{w}^{l}$ by the operators 
$\hat{l}_{\mathrm{sgn(l)}}^{|l|}$. This means that we can also write the \emph{LG} mode in Eq.(\ref{eq:DG_state}) as $\phi_{l0}^{LG} \sim \hat{l}_{\mathrm{sgn(l)}}^{|l|} \phi_{00}$. Consequently, the \emph{DG} state in Eq.(\ref{eq:DG_state}) appears as
a linear combination of modes of the form
\begin{equation}
\Phi_{n\overline{n}}=\hat{l}_{+}^{n}\,\hat{l}_{-}^{\overline{n}}\,\phi_{00}.
\label{SM_modes}
\end{equation}
The operators $\hat{l}_{\pm}$ belong to the Lie algebra of symmetries of the free 2DLSE (\ref{eq:complex_LSE}). They commute with the Schr{\"o}dinger differential operator $L_0=i \partial/\partial \tau+H$ that defines this equation. Since $\phi_{00}$ is a solution, and thus $L_0 \phi_{00}=0$, and, on the other hand, $[\hat{l}_{\pm},L_0]=0$ it is automatically guaranteed that the $\Phi_{n\overline{n}}$ modes are also solutions of Eq. (\ref{eq:complex_LSE}) \cite{Miller2012a,Wunsche1989}. Due to the linearity of the 2DLSE, another consequence of this property is that any linear combination of $\Phi_{n\overline{n}}$ modes is also a solution. This provides another alternative prove that \emph{DG} states are solutions of Eq.(\ref{eq:complex_LSE}). In fact, these modes solve a more general problem, namely, that of
the scattering of a gaussian wave packet by an \emph{arbitrary} instantaneous
potential $V(w,\overline{w})$ acting at $\tau=0$. For these reason,
we refer to them as \emph{scattering} modes (\emph{SM}). 

We determine next the functional structure of \emph{SM} by analyzing their symmetry properties. Due to its form in terms of the complex position and momentum operators (\ref{l_operators}), the operators $\hat{l}_{\pm}$
transform under  continuous rotations  as: 
\begin{eqnarray}
\hat{l}_{+} & \overset{G(\alpha)}{\rightarrow} & \epsilon \hat{l}_{+}  \nonumber\\
\hat{l}_{-} & \overset{G(\alpha)}{\rightarrow} & \epsilon^{*}\hat{l}_{-}
\end{eqnarray}
where  $\epsilon=\exp i \alpha$ is a complex $O(2)$ rotation of angle $\alpha$. In this way, under a $O(2)$ rotation $\hat{G}(\alpha)$, \emph{SM} in Eq.(\ref{SM_modes}) transform as:
\begin{equation}
\hat{G}(\alpha)\Phi_{n\overline{n}}=\exp i\alpha (n-\overline{n})\Phi_{n\overline{n}}=\epsilon^l \Phi_{n\overline{n}}.
\end{equation}
%,
Consequently, \emph{SM} are eigenfunctions of the continuous rotation operator $\hat{G}(\alpha)$ with eigenvalue $\epsilon^l$. Therefore, $l=n-\overline{n}$ is the
OAM of the \emph{SM}. In terms of their OAM, we can rearrange operators in Eq.(\ref{SM_modes}), so that \emph{SM} can also be written
as:
\begin{equation}
\Phi_{lp}=\hat{l}_{\mathrm{sgn(l)}}^{|l|}\,\hat{\triangle}^{p}\,\phi_{00},
\label{SM_modes_lp}
\end{equation}
where $\hat{\triangle}\equiv\hat{l}_{+}\hat{l}_{-}$ is a $O(2)$-invariant
operator and $p\equiv\min(n,\overline{n})$. Since the function $\Omega_{w}^{l}\equiv\{w^{|l|}\,(l\ge0),\overline{w}^{|l|}\,(l<0)\}$ transforms under continuous rotations as $\Omega_{w}^{l} \rightarrow \epsilon^l \Omega_{w}^{l}$, i.e., exactly as $\Phi_{lp}$ does, rotational symmetry determines that the functional form of a generic \emph{SM} can be always given by: 
\begin{equation}
\Phi_{lp}(w,\overline{w},\tau)=\Omega_{w}^{l}f_{lp}(|w|^{2},\tau) \phi_{00},
\label{form_of_SM}
\end{equation}
where $f_{lp}$ is an $O(2)$-invariant function depending exclusively on the modulus of $w$ (as $\phi_{00}$.)
However, the $O(2)$-invariant functions $f_{lp}$ can be \emph{explicitly}
constructed by successive application of the operators $\hat{l}_{\pm}$
on $\phi_{00}$ according to Eq.(\ref{SM_modes_lp}). Note that, due to the form of the differential $\hat{l}_{\pm}$ operators (\ref{l_operators}), their action on the gaussian function $\phi_{00}$ (\ref{eq:fundamental_mode}) always provide products of polynomials in $w$ and $\overline{w}$ times the original function $\phi_{00}$.  According to symmetry considerations, these products can be always rearranged as in Eq.(\ref{form_of_SM}) in a systematic manner, so that an analytical procedure to obtain any function $f_{lp}$ is  established.  An accurate analysis of this construction of $f_{lp}$ functions  permits to identify a general structure for them, given by:
\begin{equation}
f_{lp}\left(|w|^{2},\tau\right)=\alpha^{|l|}\beta^{p}F_{p}^{|l|}(x),
\label{flp_functions}
\end{equation}
where $\alpha=i\tau_{R}/q(\tau)$, $\beta=\tau\tau_{R}/q(\tau)$ and
$F_{p}^{|l|}(x)$ is a polynomial of $p_{\mathrm{th}}$ order in $x=\left(q(\tau)\tau\right)^{-1}\tau_{R}|w|^{2}$:
\begin{equation}
F_{p}^{|l|}(x)=\sum_{i=0}^{p}c_{pi}^{|l|}x^{i}.
\end{equation}
All coefficients in $F_{p}^{|l|}(x)$ are known in this construction.
As an example, the value  of the coefficients for the lower order $F_{p}^{|l|}$
polynomials is given in Table\ref{tab:F_polynomials}.
In addition, $F_{0}^{|l|}(x)=1$ for all $l$. 

\begin{table}
\resizebox{1.0 \columnwidth}{!} {
  \centering 
  \begin{tabular}{|c||c|c||c|c|c||c|c|c|c|}
\cline{2-10} 
\multicolumn{1}{c|}{} & \multicolumn{2}{c||}{$p=1$} & \multicolumn{3}{c||}{$p=2$} & \multicolumn{4}{c|}{$p=3$}\tabularnewline
\hline 
$c_{pi}^{|l|}$ & $c_{10}^{|l|}$ & $c_{11}^{|l|}$ & $c_{20}^{|l|}$ & $c_{21}^{|l|}$ & $c_{22}^{|l|}$ & $c_{30}^{|l|}$ & $c_{31}^{|l|}$ & $c_{32}^{|l|}$ & $c_{33}^{|l|}$\tabularnewline
\hline 
$l=0$ & 1 & -1 & 2  & -4 & 1 & 6 & -18 & 9 & -1 \tabularnewline
\hline 
$l=\pm1$ & 2 & -1 & 6 & -6 & 1 & 24 & -36 & 12 & -1 \tabularnewline
\hline 
$l=\pm2$ & 3 & -1 & 12 & -8 & 1 & 60 & -60 & 15 & -1 \tabularnewline
\hline 
$l=\pm3$ & 4 & -1 & 20 & -10 & 1 & 120 & -90 & 18 & -1\tabularnewline
\hline 
$l=\pm4$ & 5 & -1 & 30 & -12 & 1 & 210 & -126 & 21 & -1 \tabularnewline
\hline 
\end{tabular}
}
  \caption{Coefficients of the lower order $F_{p}^{|l|}$ polynomials.}
  \label{tab:F_polynomials}
\end{table}
Taking into account that we can write $\phi_{l0}^{LG} \sim \hat{l}_{\mathrm{sgn(l)}}^{|l|} \phi_{00}$, it is immediate to obtain an analytical expression for a $\left|DG(l,N)\right\rangle _{v}$ state in terms of \emph{SM} using the definition of \emph{SM} (\ref{SM_modes_lp}) in Eq.(\ref{eq:DG_state}):
\begin{equation}
\phi_{_{lNv}}^{DG}\sim\begin{cases}
\Phi_{l0}+iv\Phi_{l+N,0}+iv\Phi_{l-N,\overline{N}} & l\ge0\\
\Phi_{l0}+iv\Phi_{l+N,\overline{N}}+iv\Phi_{l-N,0} & l\le0,
\end{cases}
\label{eq:DGS_in_terms_of_SM}
\end{equation}
where $\overline{N}=\min(|l|,N)$. Because of the univocal relationship
between the unfolded and folded values $l$ and $(m,k)$, an alternative
notation for a \emph{DG} state is $\left|DG(l,N)\right\rangle _{v}=\left|DG(m,k,N)\right\rangle _{v}$.
As we will see next, the singularity structure of a \emph{DG} state crucially
depends on the value of its ``folding'' parameter $k$. 

Since \emph{SM} are found analytically so are \emph{DG} states. This property can be made explicit by substituting the expression for \emph{SM}  (\ref{form_of_SM}) into Eq. (\ref{eq:DGS_in_terms_of_SM}). However, as just mentioned, the role of $k$ is essential, so that we want to transform the conditions for $l$ in Eq.(\ref{eq:DGS_in_terms_of_SM}) in conditions for $k$. We need to distinguish between the $k = 0$ and $k \ne 0$ case. In the latter case ($k \ne 0$), which is the one we will study first, there is a biunivocal relation between the values of $k$ and $l$. Due to the fact that $l=m+kN$ and  $|m| \le N/2$, it is easy to check that the conditions $l \ge 0$ and $k \ge 1$ are equivalent provided $k \ne 0$. In the same way, it is proven that $l \le 0$ is equivalent to $k \le -1$. 
For this reason and following Eq.(\ref{eq:DGS_in_terms_of_SM}), we distinguish the $k\ge1$ ($l\ge0$) and $k\le-1$ ($l\le0$) cases in our first analysis for $k \ne 0$. 

We start by considering that $k\ge1$.  We use a symmetry argument to find the general structure of \emph{DG} states. This is the counterpart of the argument we used to determine the structure of \emph{SM} (\ref{form_of_SM})  by means of their transformation properties under $O(2)$. Now,  the transformation property of \emph{DG} functions under the  $\cal{C}_N$ symmetry   $\phi_{mkNv}^{DG} \rightarrow \epsilon_N^m \phi_{mkNv}^{DG}$ tells us that they have to be proportional  to the $\Omega_{w}^{m}$ function (which transforms as $\Omega_{w}^{m} \rightarrow \epsilon_N^m \Omega_{w}^{m} $) times a $\cal{C}_N$-invariant function. We write then $\phi_{mkNv}^{DG}$ in the following way:
\begin{eqnarray}
\phi_{mkNv}^{DG}\left(w,\overline{w},\tau\right)\ & = & \nonumber \\
& & \hspace{-7em} 
\Omega_{w}^{m}w^{\left(k-1\right)N}\mathcal{F}_{mkNv}\left(w^{N},\overline{w}^{N},|w|^{2},\tau\right)\phi_{00},
\label{eq:DG_solution_F}
\end{eqnarray}
where $ $$\mathcal{F}_{mkNv}$ is an explicitly invariant  $\cal{C}_N$ function, in the same way as $w^{\left(k-1\right)N}$ and $\phi_{00}$. Th e comparison of the result obtained after substituting  Eq.(\ref{form_of_SM}) into Eq.(\ref{eq:DGS_in_terms_of_SM}) with the general expression for the \emph{DG} state (\ref{eq:DG_solution_F}), provides us with an explicit construction for the  $\mathcal{F}_{mkNv}$ functions 
in terms of the analytical  $\mathcal{C}_{N}$-invariant functions $f_{lp}$ found previously. 

Thus, for $k \ge 1$ ($l \ge 0$) we have
\begin{equation}
\mathcal{F}_{mkNv}=\begin{cases}
f_{m+kN,0}w^{N}+ivf_{m+\left(k+1\right)N,0}w^{2N}+\\
\,\,\, ivf_{m+\left(k-1\right)N,N} &  \\
& \hspace{-4cm} m \ge 0 \,\,\,(l=l_1) \\
\\
|w|^{-2|m|}\left(f_{m+kN,0}w^{N}+ivf_{m+\left(k+1\right)N,0}w^{2N}+\right.\\
\,\,\,\left.+ivd_{mk}f_{m+\left(k-1\right)N,l}\right) & \\
& \hspace{-4cm} m \le 0 \,\,\,(l=l_2) ,
\end{cases}
\label{eq:F_functions}
\end{equation}
where $d_{mk}=|w|^{2|m|}$ if $k=1$ and $1$ if $k\ge2$. We see that for a given value of $k \ge 1$ and $|m|$, there are two possible values for $l$, $l_1=|m|+|k|N$ and $l_2=-|m|+|k|N$, depending on whether $m=|m|$ or $m=-|m|$. Since $|m| \le N/2$, these two values are positive and verify that $l_1,l_2>N$, except when $k = 1$, in which $l_2<N$. Note that, due to this fact, $\overline{N}=\min(|l|,N)$ equals $N$ in the former case and $|l_2|$ in the latter. As we will see next, these different values give rise to different types of solutions.

For $k \le-1$, one can still use the previous expressions by invoking an important $w\leftrightarrow\overline{w}$
duality symmetry of \emph{DG} states. Under the exchange between $w$ and $\overline{w}$, the fundamental gaussian function $\phi_{00}$ (\ref{eq:fundamental_mode}) is invariant. On the other hand, it is immediate to see from their definition (\ref{l_operators}) that this transformation exchange the $\hat{l}_{\pm}$ operators: $\hat{l}_{+} \stackrel{w\leftrightarrow\overline{w}}{\rightarrow} \hat{l}_{-} $.
For this reason, the \emph{LG} functions with $p=0$  ($\phi^{LG}_{l0} \sim \hat{l}_{\mathrm{sgn(l)}}^{|l|} \phi_{00}$) change under this duality transformation between $w$ and $\overline{w}$ simply as: $\phi^{LG}_{l0} \stackrel{w\leftrightarrow\overline{w}}{\rightarrow} \phi^{LG}_{-l0}$. Since the discrete deformation operator (\ref{discrete_deformation_operator}) is invariant under the exchange of $\hat{l}_{+}$ and $\hat{l}_{-}$,  the very definition of \emph{DG} states  (\ref{braket_deformation_equation}) imply that \emph{DG} states transform as \emph{LG} modes under  the duality transformation $w\leftrightarrow\overline{w}$: 
\begin{equation}
\phi_{-l,N,v}^{DG}\left(w,\overline{w},\tau\right)=\phi_{l,N,v}^{DG}\left(\overline{w},w,\tau\right),
\label{DG_duality_l}
\end{equation}
or, equivalently,
\begin{equation}
\phi_{-m,-k,N,v}^{DG}\left(w,\overline{w},\tau\right)=\phi_{m,k,N,v}^{DG}\left(\overline{w},w,\tau\right).
\label{DG_duality_mk}
\end{equation}
In terms of  the $\mathcal{F}_{mkNv}$ functions, this property reads,
\begin{equation}
\mathcal{F}_{-l,Nv}\left(w,\overline{w},\tau\right)=\mathcal{F}_{l,N,v}\left(\overline{w},w,\tau\right),
\label{F_duality_l}
\end{equation}
or, in terms of $(m,k)$,
\begin{equation}
\mathcal{F}_{-m,-k,N,v}\left(w,\overline{w},\tau\right)=\mathcal{F}_{m,k,N,v}\left(\overline{w},w,\tau\right).
\label{F_duality_mk}
\end{equation}

For $k \le -1$ ($l \le 0$), by applying this duality symmetry to Eq.(\ref{eq:F_functions}), we have
\begin{equation}
\mathcal{F}_{mkNv}=
\begin{cases}
|\overline{w}|^{-2|m|}\left(f_{-m+|k|N,0}\overline{w}^{N}+ivf_{-m+\left(|k|+1\right)N,0}\overline{w}^{2N}+\right.\\
\,\,\,\left.+ivd_{m|k|}f_{-m+\left(|k|-1\right)N,|l|}\right) & \\
& \hspace{-4cm} m \ge 0 \,\,\,(l=-l_2) ,\\
\\
f_{|m|+|k|N,0}\overline{w}^{N}+ivf_{|m|+\left(|k|+1\right)N,0}\overline{w}^{2N}+\\
\,\,\, ivf_{|m|+\left(|k|-1\right)N,N} &  \\
& \hspace{-4cm} m \le 0 \,\,\,(l=-l_1).
\end{cases}
\label{eq:F_functions_2}
\end{equation}
In this way, we have extended the expression of the  $\mathcal{F}_{mkNv}$ functions to all non-zero values of $k$. 

However, for $k \le -1$ the form of the \emph{DG} functions also changes according to  Eq.(\ref{DG_duality_mk}). Instead of Eq.(\ref{eq:DG_solution_F}), we have
\begin{eqnarray}
\phi_{mkNv}^{DG}\left(w,\overline{w},\tau\right)\ & = & \nonumber \\
& & \hspace{-7em} 
\Omega_{w}^{m}\overline{w}^{\left(|k|-1\right)N}\mathcal{F}_{mkNv}\left(w^{N},\overline{w}^{N},|w|^{2},\tau\right)\phi_{00},
\label{eq:DG_solution_F_2}
\end{eqnarray}
where  $\mathcal{F}_{mkNv}$  is now the extended function defined for all $k \ne 0$.
 
For $k=0$, we have that $l=m$ and the previous construction has to be changed accordingly. Although the symmetry arguments used to construct the \emph{DG} state functions still hold, we need to define a new set of $\cal{C}_N$-invariant functions. Therefore, instead of Eqs.(\ref{eq:DG_solution_F}) and (\ref{eq:DG_solution_F_2}), we have
\begin{equation}
\phi_{m0Nv}^{DG}\left(w,\overline{w},\tau\right)=\Omega_{w}^{m}\mathcal{G}_{mNv}\left(w^{N},\overline{w}^{N},|w|^{2},\tau\right)\phi_{00},
\label{eq:DG_solution_G}
\end{equation}
where the function $\mathcal{G}_{mNv}$ have a similar, but not identical,
structure than that of $\mathcal{F}_{mkNv}$ in terms of the $f_{lp}$
functions:
\begin{equation}
\mathcal{G}_{mNv}=\begin{cases}
f_{m0}+ivf_{m+N,0}w^{N}+iv|w|^{-2m}f_{N-m,N-m}\overline{w}^{N}\\
 & \hspace{-4cm}l=m\ge0\\
 \\
f_{|m|0}+ivf_{|m|+N,0}\overline{w}^{N}+iv|w|^{-2|m|}f_{N-|m|,N-|m|}w^{N} .& \\
 & \hspace{-4cm}l=m\le0
\end{cases}\label{eq:G_function}
\end{equation}

The final property of \emph{DG} states we want to deal in this section is biorthogonality. As  complex-argument \emph{LG} modes,  \emph{DG} states are not orthogonal but \emph{biorthogonal}. The complex-argument \emph{LG} modes $\left|LG(l,p)\right\rangle$ form a biorthogonal set, which means that there exist a \emph{different} set of states, known as adjoint states  $\left|\overline{LG}(l',p')\right\rangle$, which satisfy $\langle \overline{LG}(l',p') \left|LG(l,p)\right\rangle=\delta_{ll'}\delta_{pp'}$ \cite{Takenaka1985a}. By defining a discrete deformation of these states using the operator $\hat{D}_{v}$ analogously as we did in Eq.(\ref{discrete_deformation_operator}):
 \begin{equation}
\left| \overline{DG}(l',p',N)\right\rangle _{v}=\hat{D}_{v}(N)\left| \overline{LG}(l',p')\right\rangle,
\label{braket_deformation_equation_2}
\end{equation}
 we immediately realize that the $\{ DG(l,p,N), \overline{DG}(l',p',N) \}$ set is also biorthogonal. Indeed, inasmuch as the deformation operator is unitary, as proven in the previous section, it is true that $\hat{D}_{v}^{\dagger}\hat{D}_{v}=1$, and therefore the scalar product is preserved, so that:
 \begin{eqnarray}
\langle \overline{DG}(l',p',N)_v \left|DG(l,p,N)_v\right\rangle & = &    \nonumber \\
&   & \hspace{-3cm} \langle \overline{LG}(l',p') \left|LG(l,p)\right\rangle = \delta_{ll'}\delta_{pp'}.
\label{biorthogonality}
\end{eqnarray}

\section{Focussing dark beams}

As already mentioned, the singularity structure of \emph{LG} modes is simple. In the case that the mode carries OAM, i.e., when $l \ne 0$, there exists a single phase singularity located at the symmetry axis of the mode. The topological charge of this singularity coincides with the OAM of the mode $q=l$. In a representation of the field amplitude, the trajectory followed by this singularity in the $xy\tau$ space is just a straight line. This straight line constitutes a dark ray where the intensity of the field vanishes. A characteristic example of such a dark ray is shown in Fig.~\ref{fig1}(a), where we can see a view of the mode from two different view points. We also observe in this figure that the amplitude profile of the mode hosting this dark ray exhibit at the same time the perfect $O(2)$ symmetry of \emph{LG} modes. It is clear that the action of the discrete deformation operator changes the continuous rotational properties of the \emph{LG} mode by transforming it into a \emph{DG} state with discrete rotational symmetry. However, as Figs.~\ref{fig1}(b) and (c) unveil, this transformation can occur in two completely different ways. We can see that two different discrete deformations of the same \emph{LG} mode ---Fig.\ref{fig1}(a)--- can produce either a simple modulation of the amplitude without changing the original dark ray  ---Fig.\ref{fig1}(b)--- or give rise to a completely new multi singular structure formed by a bundle of rays converging at the waist plane ---Fig.\ref{fig1}(c). We shall refer to this bundle of converging dark rays as the \emph{focussing dark beam} associated to the \emph{DG}  state. We will show next how the previous mechanism of generation of focussing dark beams is linked to the properties of the \emph{DG} states analyzed in the previous section. In particular, we will learn the key role played by the folding parameter $k$ to understand the generation or not of dark beams embedded in a given \emph{DG} state.

The singularity structure of \emph{DG} states arises from the condition $\phi_{mkNv}^{DG}=0$. According to Eqs.(\ref{eq:DG_solution_F}), (\ref{eq:DG_solution_F_2}) and  (\ref{eq:DG_solution_G}), the zeroes of $\phi_{mkNv}^{DG}$ occur when one of the following conditions are met:
\begin{enumerate}
  \item $\Omega_{w}^{m}w^{\left(k-1\right)N}=0$ ($k\ge1$),
  \item $\Omega_{w}^{m}\overline{w}^{\left(|k|-1\right)N}=0$ ($k\le-1$),
  \item $\mathcal{F}_{mkNv}$=0 ($k \ne 0$), or
  \item $\Omega_{w}^{m}=0$ ($k=0$),
  \item $\mathcal{G}_{mNv}=0$ ($k=0$).
\end{enumerate}
 In the quest of singularities of \emph{DG} states, it is important to know what occurs with singularities on axis. In this context, an important property of the  $\mathcal{F}_{mkNv}$ and  $\mathcal{G}_{mNv}$  functions  is that they do not show zeros when $w=0$ and $\tau \ne 0$ (provided $|m| < N/2$; the case $|m|=N/2$ should be analyzed separately.) This property can be proven by taken the limit $w \rightarrow 0$ in Eqs.(\ref{eq:F_functions}), (\ref{eq:F_functions_2}) and (\ref{eq:G_function}). In all cases, both functions tend to a quantity proportional to some $f_{lp}$ function evaluated at $w=0$, being its corresponding proportionality constant always different from zero if $\tau \ne 0$. However, according to the form of $f_{lp}$ functions (\ref{flp_functions}), when $\tau \ne 0$ $f_{lp}(0)$ is non-vanishing since the polynomial $F_{p}^{|l|}(x)$ always shows non-zero values for its zero order terms. 

The fact that both $\mathcal{F}_{mkNv}$
and $\mathcal{G}_{mNv}$ have no zeroes at $w=0$ determines that
axial singularities at $w=0$ and $\tau\ne0$ are given by one of
the three previously presented conditions: (i) or (ii), if $k \ne 0$; or, alternatively, (iv) if $k=0$. A simple analysis of these expressions permit to establish that a generic $\left|DG(m,k,N)\right\rangle _{v}=\left|DG(l,N)\right\rangle _{v}$ state with $l \ne 0$ necessarily presents a singularity located at the axis $w=0$ with topological charge: 
\begin{equation}
q_{\mathrm{ax}}=m+\mathrm{sgn}(k)\left(|k|-1\right)N=l-\mathrm{sgn}(k)N\,\,\,(\tau\ne0).
\label{eq:axial_topological_charge}
\end{equation}
On the other hand, the value of the topological charge at the waist ($\tau = 0$) is always $l$. This is so due to the unitary nature of the deformation operator. According to the modified waist condition  (\ref{waist_condition_discrete}), the action of this operator at $\tau =0$ on the \emph{LG} mode  is simply a multiplication by the unimodular complex function $\exp i V$ (recall $V$ is a real function.) Inasmuch as  $\exp i V$ cannot be zero, the zero of the  \emph{DG} state  at $\tau=0$ is the same as that of the \emph{LG} mode, i.e., it is located at $w=0$ and it has topological charge $q_{\mathrm{ax}}=l$. 

We immediately recognize an important qualitative difference in the $q_{\mathrm{ax}}(\tau)$ function when comparing the $k \ne 0$ and $k = 0$ cases. For $k=0$, Eq.(\ref{eq:axial_topological_charge}) tells us that the axial charge function  $q_{\mathrm{ax}}$ is continuous for all values of the evolution parameter $\tau$. Moreover, it is a constant function that takes always the value $q_{\mathrm{ax}}=l$, exactly as occurs in the \emph{LG} mode from which it is derived. However, when $k \ne 0$, the axial charge function experiments two qualitative changes: firstly, it develops a discontinuity at $\tau = 0$, and, secondly, its value for $\tau \ne 0$ is no longer $l$ but $l-\mathrm{sgn}(k)N$. We can understand now better the results already presented in Fig.~\ref{fig1}, which provide a neat visualization of this analysis. We see that the \emph{DG} state in Fig.~\ref{fig1}(b) is a $k = 0$ state. Consequently, in agreement with our previous argument, the topological charge function is continuous and constant and it physically corresponds to a single dark ray with charge $q_{\mathrm{ax}}=l$, identical to the one of the \emph{LG} mode in Fig.~\ref{fig1}(a). In Fig.\ref{fig1}(c) we present a discrete deformation of the the same \emph{LG} mode in Fig.~\ref{fig1}(a), but now with $k \ne 0$. We see that the topological charge at $\tau =0$ is still $l$. However, for the rest of values of $\tau$ $q_{\mathrm{ax}}=l-N$. The physical process associated to this discontinuity in the axial charge is clearly visualized in Fig.~\ref{fig1}(c). We see how this discontinuity is produced by the presence of $N$ off-axis singularities focussing at $\tau = 0$ and symmetrically distributed around the symmetry axis. So, the discontinuity in the $q_{\mathrm{ax}}(\tau)$ function is intimately related to the generation of a \emph{focussing} dark beam. Since the discontinuity in the axial charge function occurs only for $k \ne 0$ \emph{DG} solutions, we have here a clear signal that the generation of a focussing dark beam is determined by the non-zero value of the unfolding parameter $k$.

We can rigorously prove our last statement  by analyzing the $w\rightarrow0$ and $\tau\rightarrow0$ limits of
the $\mathcal{F}_{mkNv}=0$ and $\mathcal{G}_{mNv}=0$ conditions. In this way, we can unveil the off-axis singularity structure of a given \emph{DG} state. We first go  to Eqs.(\ref{eq:F_functions}) and (\ref{eq:F_functions_2}) and find the form of $\mathcal{F}_{mkNv}$ functions by taking into account that in this regime we can neglect the $O(w^{2N})$ terms and that $f_{lp}$ functions in Eq.(\ref{flp_functions}) can be approximated as $f_{lp}\sim (-i \tau)^p F^{|l|}_p(0)$. 

For $k \ge 1$ ($l \ge 0$), we find that a for a given \emph{DG} state characterized by the indices $(l,N) \Leftrightarrow (m,k,N)$, the $\mathcal{F}_{mkNv}=0$ condition becomes near the origin:
\begin{eqnarray}
 \lefteqn{w^N + i v (-i \tau)^N \gamma_{lN} \approx 0,}  \nonumber \\
 &  & \hspace{2cm} (\text{for } l=l_1 \, \text{or }  l=l_2 \,\,\text{with } k \ge 2) \Leftrightarrow l>N  \nonumber \\
&  &  \nonumber \\
   \lefteqn{w^N + i v |w|^{2|m|} (-i \tau)^{l} \gamma'_{lN} \approx  0,}  \nonumber \\ 
& &  \hspace{2cm}   (\text{for } l=l_2 \,\,\text{with } k = 1) \Leftrightarrow 0<l<N,
\label{zeroes_F}
\end{eqnarray}
where $\gamma_{lN}\equiv F^{|l|-N}_N(0)$ and $\gamma'_{lN}\equiv F^{|l|-N}_l(0)$. This property shows that, indeed, $N$ off-axis zeroes of $\mathcal{F}_{mkNv}$
occur at the same radial position $r_0=|w_0|$ given by:
\begin{eqnarray}
r_0 & = &   (-i v)^{1/N} \gamma_{lN}^{1/N}  (-i \tau) \nonumber \\
&  &   \hspace{1cm}   (\text{for } l=l_1 \, \text{or }  l=l_2 \,\,\text{with } k \ge 2) \Leftrightarrow l>N   \nonumber \\
r_0 & = &   (-i v)^{1/N} \gamma_{lN}^{\prime 1/N} (-i \tau)^{\frac{l}{2l-N}}. \nonumber \\
& &     \hspace{1cm} (\text{for } l=l_2 \,\,\text{with } k = 1) \Leftrightarrow 0<l<N.
\label{radial_positions}
\end{eqnarray}
Both type of phase singularities tend to zero in the $\tau \rightarrow 0$ limit. In the second case, let us emphasize that the exponent of $\tau$ in this expression is always finite and positive since $l=l_2$>0 and $2 l_2 - N= N-2|m|>0$  because we are excluding explicitly the $|m|=N/2$ case and, therefore, our constraint on $m$ is  $|m|<N/2$. So that, there is no singularity at $\tau=0$. Consequently, we have shown that there exist $N$ singularities with charge $q=+1$ approaching symmetrically to the axis when $\tau\rightarrow0$. For $k\le-1$ ($l \le 0$), we would obtain an equivalent property but for $q=-1$ charges corresponding
to the $w\leftrightarrow\overline{w}$ duality symmetry of the $\mathcal{F}_{mkNv}$ functions. The two different behaviors in Eqs.(\ref{zeroes_F}) and (\ref{radial_positions}) would  correspond then to $|l|>N$ in the first case and to $|l|<N$ in the second. In all cases we conclude that  any \emph{DG} state with $k \ne 0$ will generate a focussing dark beam.

However, the situation is completely different for $k=0$ since $\mathcal{G}_{mNv}$
does not show any zero approaching $w=0$ when $\tau\rightarrow0$. We can see this property by writing the condition $\mathcal{G}_{mNv}=0$ close to the origin in a similar way as we did before ---we use now Eq.(\ref{eq:G_function}). We have:
\begin{eqnarray}
 1+  i v |w|^{-2|m|} (-i \tau)^N \gamma_{mN} \overline{w}^N  & \approx & 0, \hspace{1cm} (m \ge 0)  \nonumber \\
 &  &   \nonumber \\
 1+  i v |w|^{-2|m|} (-i \tau)^N \gamma_{mN} w^N  & \approx & 0,  \hspace{1cm} (m \le 0) \nonumber 
\end{eqnarray}
The zeroes of these complex equations are located at the radial position $r_0 \sim \tau^{-|m|/(N-2|m|)}$, so that they diverge when $\tau \rightarrow 0$ since $|m|<N/2$. Hence, the zeroes of   $\mathcal{G}_{mNv}$ cannot connect to the extant singularity at $w=0$ in the $\tau \rightarrow 0$ limit..

In summary, we have demonstrated that \emph{DG} states with $k=0$ do not show any bifurcation at $\tau=0$.
On the contrary, all \emph{DG} states with $k\ne0$ exhibit a \emph{focussing}
point at $(w,\tau)=(0,0)$, where all trajectories of phase singularities
converge. Therefore, a \emph{focussing} dark beam structure is embedded
in a \emph{DG} state \emph{if and only if} $k\ne0$ ---see Fig.\ref{fig1}(c).
For $k=0$, a single axial dark ray is present when $l\ne0$ and only
a modulation in the amplitude reveals the discrete nature of the \emph{DG}
state as compared to a \emph{LG} mode with the same value of $l$ ---compare
Figs.\ref{fig1} (a) and (b). Because all dark rays of a \emph{DG}
state with $k\ne0$ focus at $\tau=0$, we call the $w=0$ point in
this plane the \emph{dark focus }of the \emph{DG} state. 
\begin{figure}
\includegraphics[width=1\columnwidth]{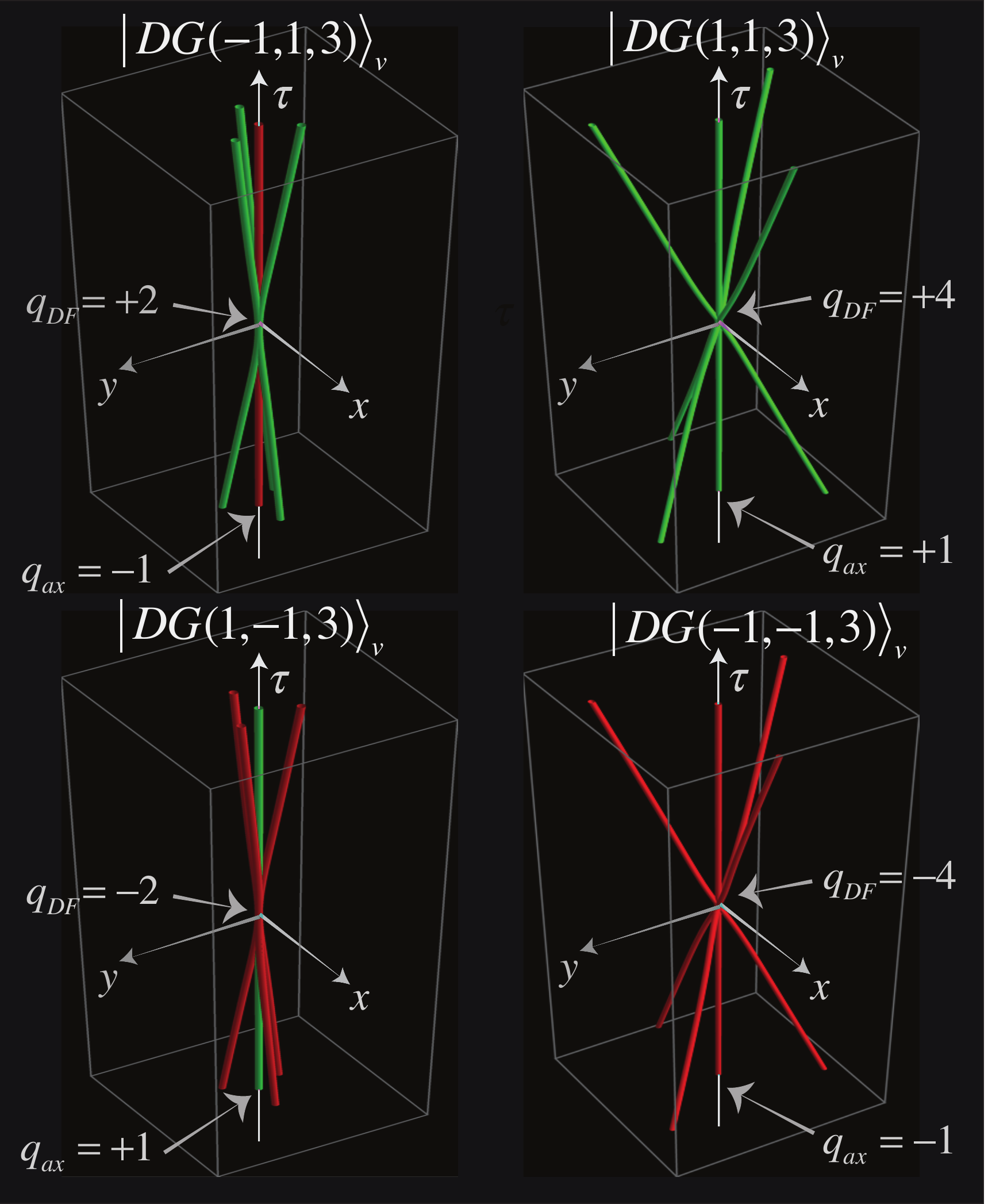}
\caption{Structure of focussing dark beams embedded in lower order
 states near the \emph{dark focus} for a quadruplet with $|k|=1$ and $|m|=1$ for $N=3$.   The two states on the left column correspond to $l =\pm l_2=\pm2$ whereas the ones in the right column correspond to $l=\pm l_1=\pm 4$. {[}$\tau_{R}=1$;
$v=0.1$; $\tau_{\mathrm{min}}=-3$; $\tau_{\mathrm{max}}=3$; transverse
range: left $L=0.75$, right $L=1.5${]}.
\label{fig2}}
\end{figure}

The paradigmatic structure of characteristic focussing dark beams corresponding to \emph{DG} states with $k \ne 0$ are given in Fig.\ref{fig2}. For a given value of $|m|$ and $|k|$ there are four possible \emph{DG} states generated by combining the signs of $m= \pm |m|$ and $k= \pm |k|$. As we have seen in the previous section, these four states are related by the $w\leftrightarrow\overline{w}$ duality symmetry ---see Eqs.(\ref{DG_duality_l}) and (\ref{DG_duality_mk}). These four states are: $\left|DG(\pm l_1,N)\right\rangle _{v}$ and $\left|DG(\pm l_2,N)\right\rangle _{v}$  ---where $l_1=|m|+|k|N$ and $l_2=-|m|+|k|N$--- or, equivalently using $(m,k)$ indices, $\left|DG(\pm |m|,\pm |k|,N)\right\rangle _{v}$ and $\left|DG(\mp |m|,\pm |k|,N)\right\rangle _{v}$. Our previous analysis of the trajectories of phase singularities close to the origin reflected in Eq.(\ref{radial_positions}) points out a qualitatively different behavior for focussing dark beams generated by the states  $\left|DG(\pm l_2,N)\right\rangle _{v}$ when $|k|=1$ as compared to their quadruplet counterparts $\left|DG(\pm l_1,N)\right\rangle _{v}$. In order to explicitly visualize this difference, we present in  Fig.\ref{fig2} a quadruplet of states corresponding to the $|k|=1$ and $|m|=1$ case, in which the pair of states $\left|DG(\pm l_2,N)\right\rangle _{v}$  fulfilling the condition $|l|<N$ are shown in the left column and the other pair $\left|DG(\pm l_2,N)\right\rangle _{v}$ fulfilling $|l|>N$ is shown in the right. According to Eq.(\ref{radial_positions}), when $|k|=1$, the $\tau$ dependence of the radial coordinate of phase singularities is different for  $\left|DG(\pm l_2,N)\right\rangle _{v}$ states, as compared to the rest of cases. This feature can be clearly appreciated in Fig.~\ref{fig2} by comparing the different behavior of dark beams near the origin in the left and right columns.
\begin{figure}
\includegraphics[width=1\columnwidth]{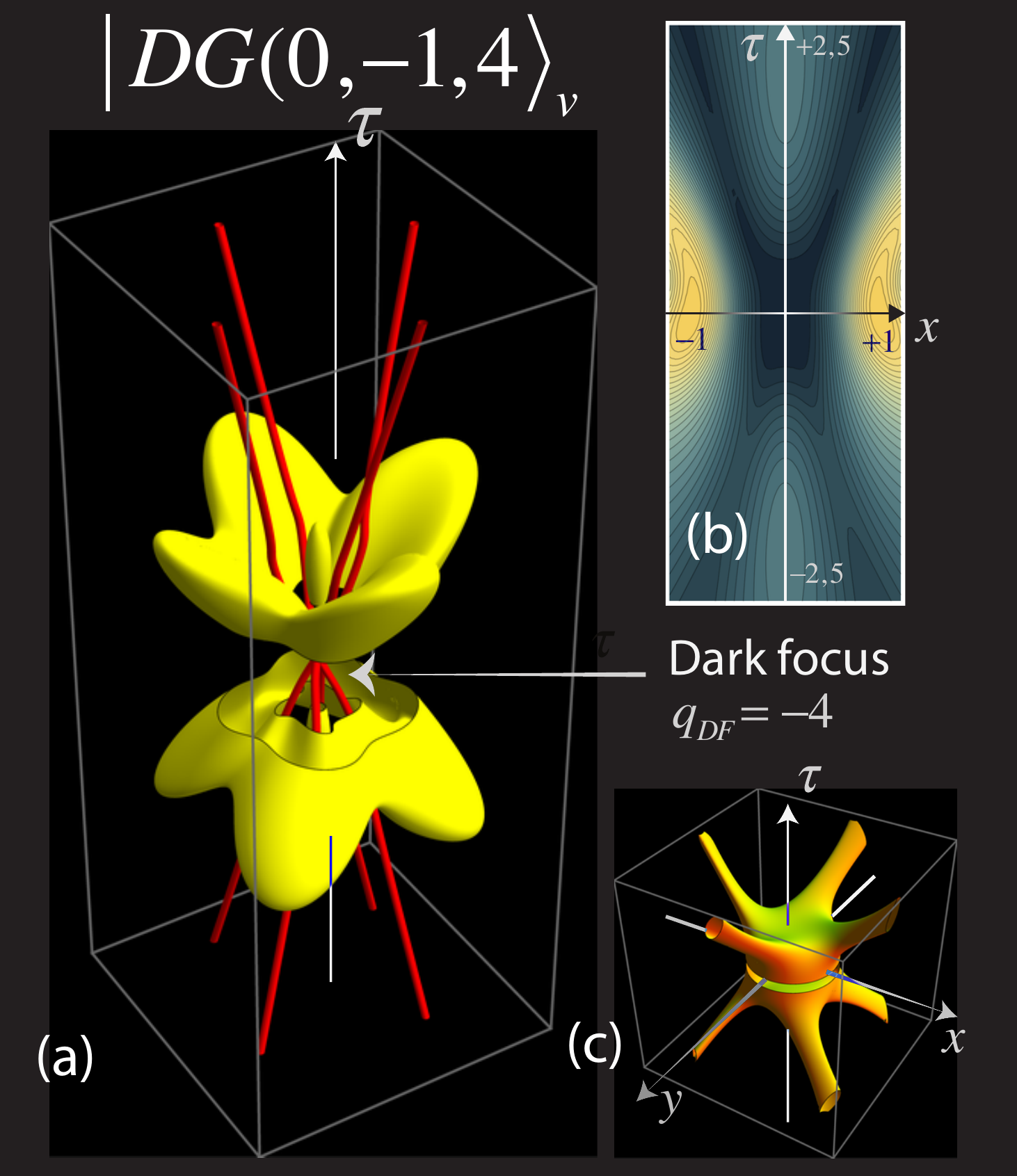}
\caption{Dark focus structure of a DG state with $m=0$ and no axial dark ray
{[}$\tau_{R}=1$; $ $$v=0.25${]}: (a) Amplitude at 1/4-maximum and
dark beam with $q_{ax}=0$ and $q_{df}=-4$; 3D box is $8\times8\times20$.
(b) Dark focus \emph{xz} section. (c) 3D representation of the dark
focus region (1/30-maximum); 3D box is $1\text{\ensuremath{\times}1\ensuremath{\times}2.5}$.
\label{fig3}}
\end{figure}

The dark focus is one of the most distinguishing features of a \emph{DG} state. A general property of the dark focus is apparent in the examples provided in Fig.~\ref{fig2}.
The axial charge, as dictated by Eq.(\ref{eq:axial_topological_charge}), is $l-\mathrm{sgn}(k)N$ for $\tau\ne0$. However, the convergence of the $N$ dark rays of the \emph{DG} state in $\tau=0$ determines the topological charge of the dark focus to be precisely $l$: $q_{DF}=l$. In this way,
Eq.(\ref{eq:axial_topological_charge}) can be understood now as a conservation law for the topological charge: $q_{DF}=q_{\mathrm{ax}}+\mathrm{sgn}(k)N$.
For $k>0$ ($k<0$), off-axis dark rays correspond to $+1$ ($-1)$ charges. The conservation of $l$, despite being no longer the OAM of the state, can be interpreted as the topological conservation law associated to the $\mathcal{C}_{N}$ discrete rotational symmetry of the \emph{DG} state.

Besides, \emph{DG}  states with $q_{DF}=\pm|l|$ present a dark beam structure that is related by the $w\leftrightarrow\overline{w}$ duality symmetry. In cartesian coordinates this symmetry is equivalent to the mirror reflection:
\begin{equation}
\mathrm{R}_{x}: (x,y)\overset{\mathrm{R}_{x}}{\rightarrow}(x,-y),
\label{mirror_reflection}
\end{equation}
together with a \emph{simultaneous}  charge conjugation $q \rightarrow -q$ of all  topological charges.
We can check this symmetry in Fig.\ref{fig2} as well. Position of dark rays for the states in the lower row can be obtained, respectively, by properly mirror
reflecting with respect the $x$ axis the dark beams of the upper
row along with charge conjugation (red/green color exchange in Fig.\ref{fig2}.)

\emph{DG} states with $k \ne 0$ exhibit a rich diversity of dark beams structures embedded in their gaussian-like amplitudes. The form of a generic solution of a \emph{DG} state, such as given in Eqs (\ref{eq:DG_solution_F})  and  Eqs (\ref{eq:DG_solution_F_2}), indicates that the properties of the dark beam, encoded in the $\mathcal{F}_{mkNv}$ function, and of the bright part of the beam, encoded in the gaussian function $\phi_{00}$, present a certain degree of independence. We see that the $v$ parameter only affects the dark beam function $\mathcal{F}_{mkNv}$ function, whereas $\tau_R$ and the beam parameter $q(\tau)$ appear in both functions, but in completely different functional ways. Thus we expect some type of interplay between dark beams and bright amplitudes in terms of these parameters. In Fig.~\ref{fig3} we present an interesting
case of \emph{DG} state characterized by $m=0$ and $|k|=1$. In such a state, the dark focus does not exhibit a dark ray on axis since, according to Eq.(\ref{eq:axial_topological_charge}), its axial charge is $q_{\mathrm{ax}}=0$ for all $\tau \ne 0$. The only on-axis singularity is located at $\tau=0$ being absent for $\tau\ne0$. Besides, the interplay between the bright part of the beam and the dark beam here is strong. This fact is reflected in the remarkable modulation of the amplitude near the dark rays, visible in Fig.~\ref{fig3}(a). In this case, as mentioned before, a strong interplay between the bright and dark parts of the beam is achieved by increasing the value of $v$ (larger than in previous cases.) The 2D and 3D representations of the dark focus region near the origin in Figs.~\ref{fig3}(b) and \ref{fig3}(c) reveals a combination of high intensity gradients with high phase contrasts (note that the topological charge at the focus is $q_{DF}=4$.)

\begin{figure}
\includegraphics[width=1\columnwidth]{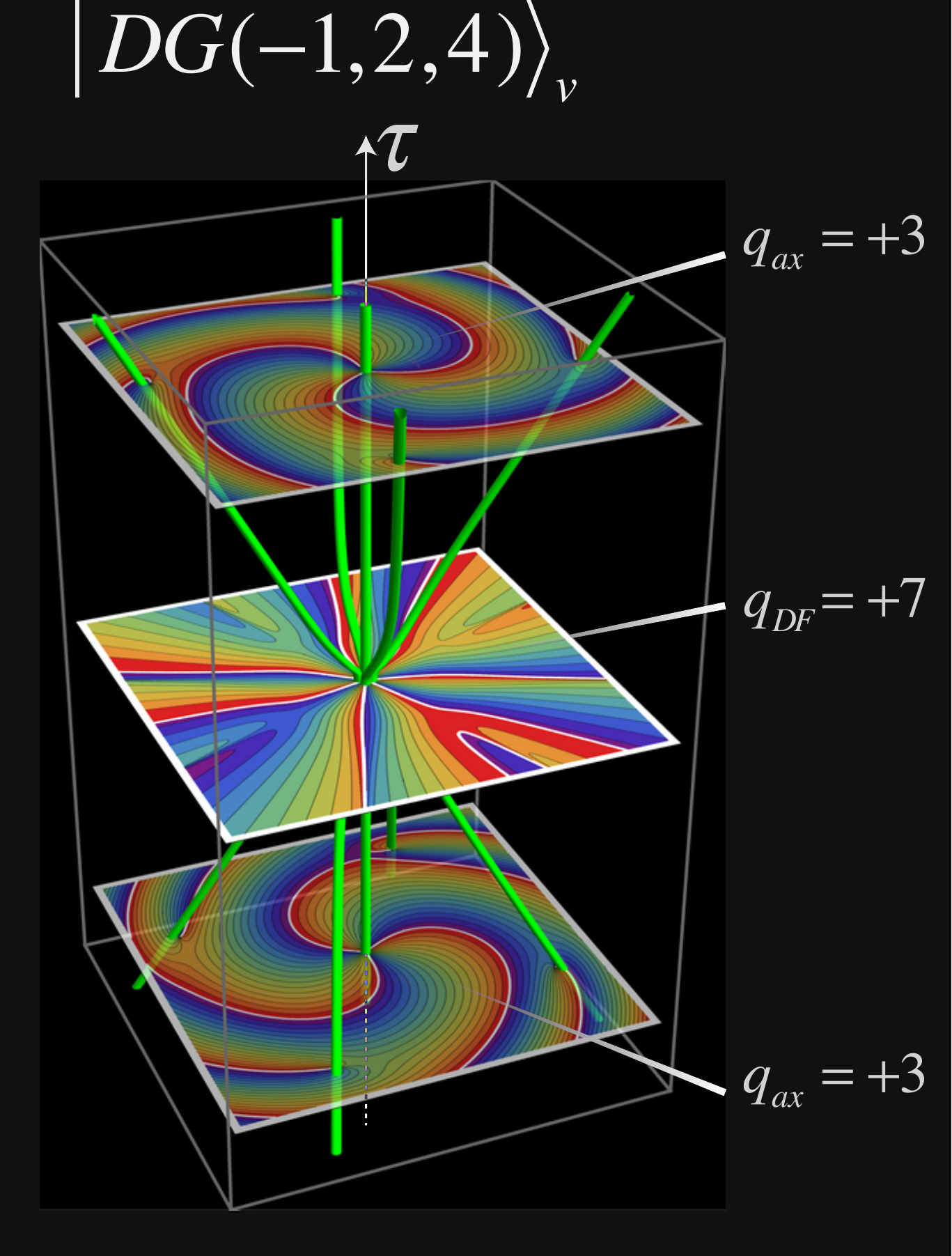}
\caption{\emph{DG} state with $k=2$ generating a focussing dark beam with a highly-charged
singularity on axis {[}$\tau_{R}=1$,$v=0.1${]}. Phase profiles are
represented at $\tau=-5$, $0$ and $+5$. 3D box is $7\times7\times12$.
\label{fig4}}
\end{figure}

In discrete symmetry media, the presence of a $\cal{C}_N$-invariant potential owning discrete rotational symmetry and extending infinitely in $\tau$ forces the axial charge of a vortex to be constrained by the rule $|q_{\mathrm{max}}| < N/2$ \cite{Garcia-March2009a,Zacares2009}. On the contrary, \emph{DG} states with highly-charged singularities on axis are allowed by the topological law (\ref{eq:axial_topological_charge}) beyond this cutoff  provided $|k|\ge 2$. In Fig.~\ref{fig4} we present an example of such a state with $k=2$. The dark beam pattern is, nevertheless, the same as for any other \emph{DG} state. $N$ single off-axis  phase singularities merge with the axial singularity at the dark focus once  and then diverge. The difference now is that the axial charge $q_{\mathrm{ax}}=3$ exceeds the maximum value for the axial charge allowed by the previous rule  for discrete potentials (in this case, $|q_{\mathrm{max}}|=1$, for $N=4$.) Note that the later rule applies to potentials that act during an infinitely long period in $\tau$, whereas \emph{DG}  states are associated to the action of instantaneous potentials. In this way, the seeming contradiction is removed.

\section{Conclusions}

Generation of \emph{DG} states is possible because the discrete deformation operator, generated by the instantaneous $\mathcal{C}_{N}$
potential, acts as a ``state converter'' changing an \emph{LG} mode for
$\tau<0$ into a \emph{DG} state for $\tau>0$. Since the form of the
potential appearing in the discrete deformation operator $\hat{D}_{v}$
is valid for general real discrete potentials for small $w$ (up to
a global rotation), approximated \emph{DG} states are expected to appear
in scattering or diffraction experiments in which $O(2)$ symmetry
is broken. Recent experiments of vortex diffraction in optics using
discrete diffractive optical elements (DOE) show, in fact, output states that can be assimilated
to \emph{DG} or quasi \emph{DG} states \cite{molina-terriza-prl87_23902a,Jack2008a,Hickmann2010b,Ferreira2011,Gao2012a,Novoa2014a}.
These experiments can be reinterpreted as examples of discrete deformation
operators generating \emph{DG} states in $\tau>0$ out of \emph{LG} modes
in $\tau<0$. 

 The general framework here presented opens the door to the control
of \emph{DG} states and dark beams beyond previously proposed strategies
\cite{ferrando2013}. It is feasible to find other type of instantaneous
potentials ---not necessarily real--- leading to different deformation
operators acting as generalized ``state converters'' between arbitrary
\emph{DG} states. As an example, it is possible to design potentials transforming
a \emph{DG} state with a dark focus on a given value of $\tau$ into a
\emph{DG} state with a dark focus in a different position. This designed
potential would act as a lens for dark rays imaging one dark focus
onto the other. Its experimental feasibility in optics is realistic
using current encoding techniques to design DOE with arbitrary phase profiles \cite{Moreno2010b}. A similar strategy
including reflecting optical elements \cite{Marcuse1982} would permit
the design of \emph{DG} resonators  acting as dark beam cavities. A novel
geometrical optics for dark rays can be then envisaged in analogy to
the classical geometrical optics used for the manipulation of ordinary bright rays \cite{ferrando2013}. 

Besides
the control of dark rays, \emph{DG} states present a rich and versatile
structure for the gradients of both the phase and amplitude of the
field. Thus, the present formalism can be of help to design adequate
optical forces for optical trapping of small neutral particles, atoms
and molecules \cite{Ashkin2000a,Grier2003a,Roichman2008a,Woerdemann2013a}. We have seen that \emph{DG} states present the possibility to manipulate their dark (i.e., phase) and bright (i.e., intensity) profiles with a certain degree of independence. This feature combined with the potential control of dark beams using DOE, which are also standard tools for manipulating gaussian beams, permits to foresee interesting applications in optical trapping. It is remarkable here that \emph{DG} states are experimentally obtained by simple diffraction using discrete DOE \cite{Gao2012a,Novoa2014a}, instead of by multiple interference of \emph{LG} modes, as other multi-singular solutions \cite{Indebetouw1993,jenkins-joa3_527a,Chavez-Cerda2001a,Bandres2004,Bandres2004a,Volyar2006,Deng2008a,Gutierrez-Vega2008a,Fadeyeva2012a,Steuernagel2012a,Dorilian2013a,Martinez-Castellanos2013a}.

Finally, it should not be ignored the potential application of \emph{DG} states
in quantum optics and quantum mechanics, which is based in the fact
that \emph{DG} states form a biorthogonal set in the same way as the complex-argument
\emph{LG} modes they are derived from \cite{Takenaka1985a}. So, \emph{DG} states can be legitimately used as a basis
for operator expansions of the quantum field in the same way as plane waves (momentum expansion) or \emph{LG} modes (angular momentum expansion). The advantage here is that they present a richer phase singularity structure than other gaussian modes. Besides, they provide an expansion  in a different quantum number, namely, the discrete angular momentum $m$. Quantum states based on the discrete angular momentum $m$ can provide an alternative to high-dimensional quantum spaces based on OAM \cite{Molina-Terriza2007b}. Additionally, they present a different quantum operational algebra and a more complex spatial mode structure that can bring a new perspective for quantum information processing  \cite{Krenn2013a}.

The author acknowledges support from the Spanish grant TEC2010-15327
from MINECO. 

\bibliographystyle{apsrev4-1}
\bibliography{ferrando_arXiv_v2_2014}

\end{document}